\documentclass[sigconf, nonacm]{acmart}

\usepackage{pvldb}

\renewcommand\vldbdoi{XX.XX/XXX.XX}
\renewcommand\vldbpages{XXX-XXX}
\renewcommand\vldbavailabilityurl{https://github.com/RUCKBReasoning/DiagDBench} 

\usepackage{enumitem}
\usepackage{xcolor}
\usepackage{amsmath}
\usepackage{graphicx}
\usepackage{float}
\usepackage{adjustbox}
\usepackage{algorithm}
\usepackage{algorithmic}
\usepackage{multirow}
\usepackage{tabularx}
\usepackage{booktabs}

\newcommand{\vpara}[1]{\vspace{0.05in}\noindent \textbf{#1 }}
\newcommand{\bench}{\textsc{DiagDBench}}
\newcommand{\sbench}{\textsc{DiagDBench} }

\begin{document}
\title{\bench: Evaluating Agentic LLMs on Interactive Database Diagnosis and Anomaly Remediation [Experiment, Analysis \& Benchmark]}

\author{
    Xinxin~Zhao$^{1,2,5}$,
    Rong~Kang$^5$,
    Tieying~Zhang$^5$,
    Cuiping~Li$^{1,2}$,
    Hong~Chen$^{1,3}$,
    Jing~Zhang$^{1,4\dagger}$
}
\affiliation{
    $^1$ School of Information, Renmin University of China, Beijing, China, 
    $^2$ Beijing Key Laboratory of Intelligent Database Systems and Applications, Beijing, China,
    $^3$ Engineering Research Center of Database and Business Intelligence, MOE, China, 
    $^4$ Key Laboratory of Data Engineering and Knowledge Engineering, MOE, China,
    $^5$ ByteDance Inc. \country{China} 
}

\email{
  {zhaoxinxin798, licuiping, chong, zhang-jing}@ruc.edu.cn
}
\email{
    {kangrong.cn, tieying.zhang}@bytedance.com
}

\thanks{$\dagger$ Jing Zhang is the corresponding author.}

\begin{abstract}
Database diagnosis and anomaly remediation are essential to reliable database management system (DBMS) operation. 
Despite advances in large language model (LLM)-based database agents, constructing live DBMS environments that reproduce diverse anomalies at scale remains challenging, limiting the availability of executable samples for database agent training and evaluation. 
To this end, we introduce \bench, a large-scale benchmark for training and evaluating database agents on interactive diagnosis, anomaly remediation, and optional workload performance optimization across diverse, realistic anomaly scenarios.
Each sample pairs a user request, comprising an issue description and an anomaly-verified workload under which the issue was observed, with a self-contained, executable DBMS environment that supports real-time monitoring, interactive diagnosis, remediation, and workload execution.
To construct these samples at scale, we propose an LLM-based framework with two complementary pipelines: \textit{Manual-Guided Anomaly Synthesis}, which sequentially synthesizes user requests, database modifications, and anomaly-injection scripts grounded in DBA manuals and official DBMS documentation; and \textit{Incident-Guided Anomaly Reproduction}, which reconstructs anomaly samples end to end from real-world enterprise online diagnosis incidents.
Built using this framework with sample filtering, expert review, and refinement, \sbench comprises 6,940 synthesized training samples and a held-out test set of 454 expert-annotated samples, spanning three hardware configurations, two DBMSs, and three database schemas.
Extensive evaluations of database agents powered by multiple LLMs demonstrate that even the best-performing agent, ReAct with GPT-6 Astra, scores only 56.24, whereas most agents score below 45, underscoring the difficulty of \bench.

\end{abstract}


\maketitle

\vldbtopmatter

\newcommand{\figoverview}{
\begin{figure*}[t]
    \centering
    \includegraphics[width=0.88\textwidth]{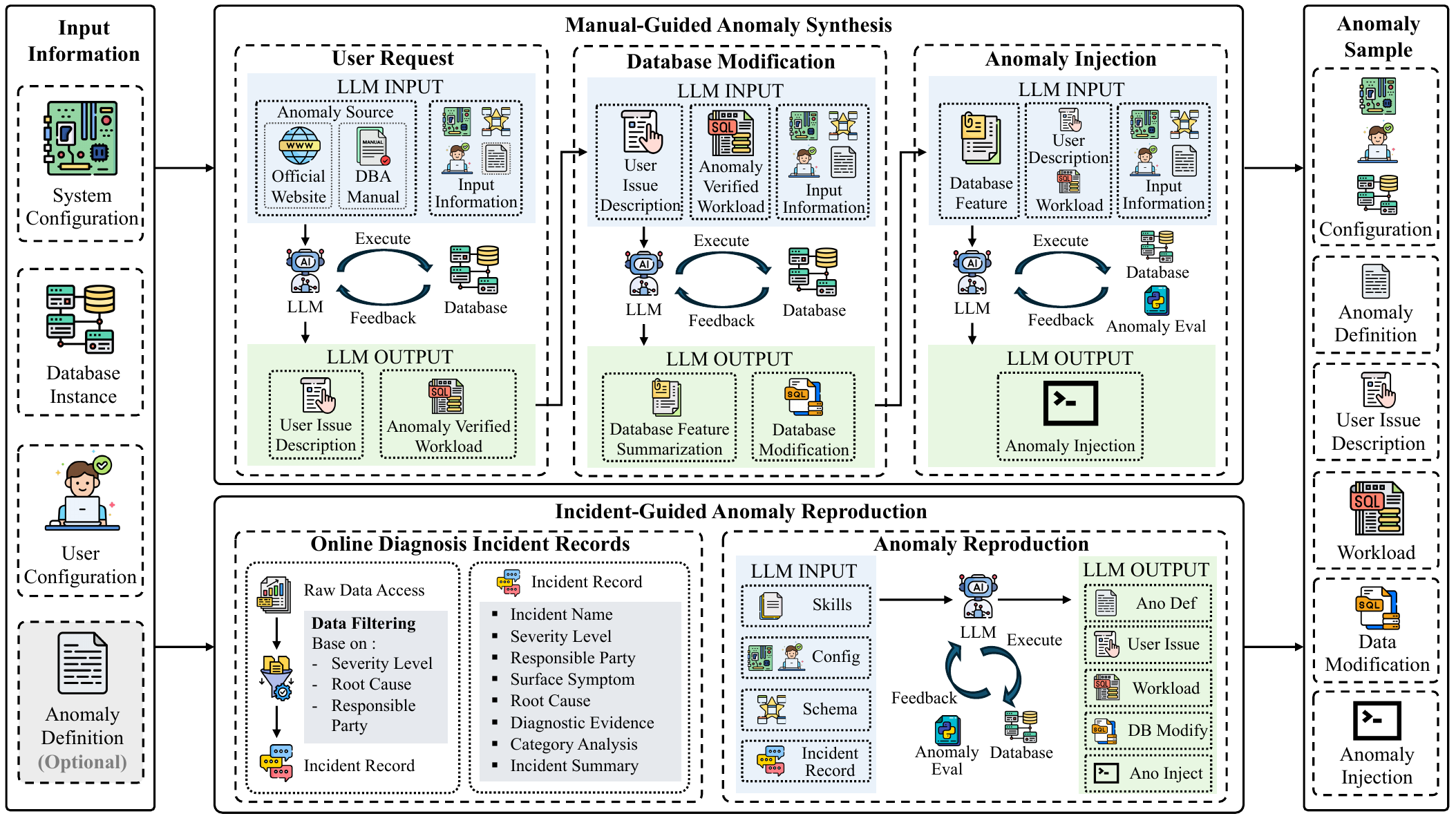}
	\caption{\label{fig:overview} Overview of the LLM-based database anomaly sample construction framework.}
\end{figure*}
}

\newcommand{\figexample}{
\begin{figure}[t]
    \centering
    \includegraphics[width=0.46\textwidth]{figures/2610171231/Example_0.png}
	\caption{\label{fig:example} Illustration of a real-world user-observed anomaly reproduced in a TPC-H database, while the monitored time-series system metrics remain within normal ranges.}
\end{figure}
}

\newcommand{\figanalysis}{
\begin{figure*}[t]
    \centering
    \includegraphics[width=0.98\textwidth]{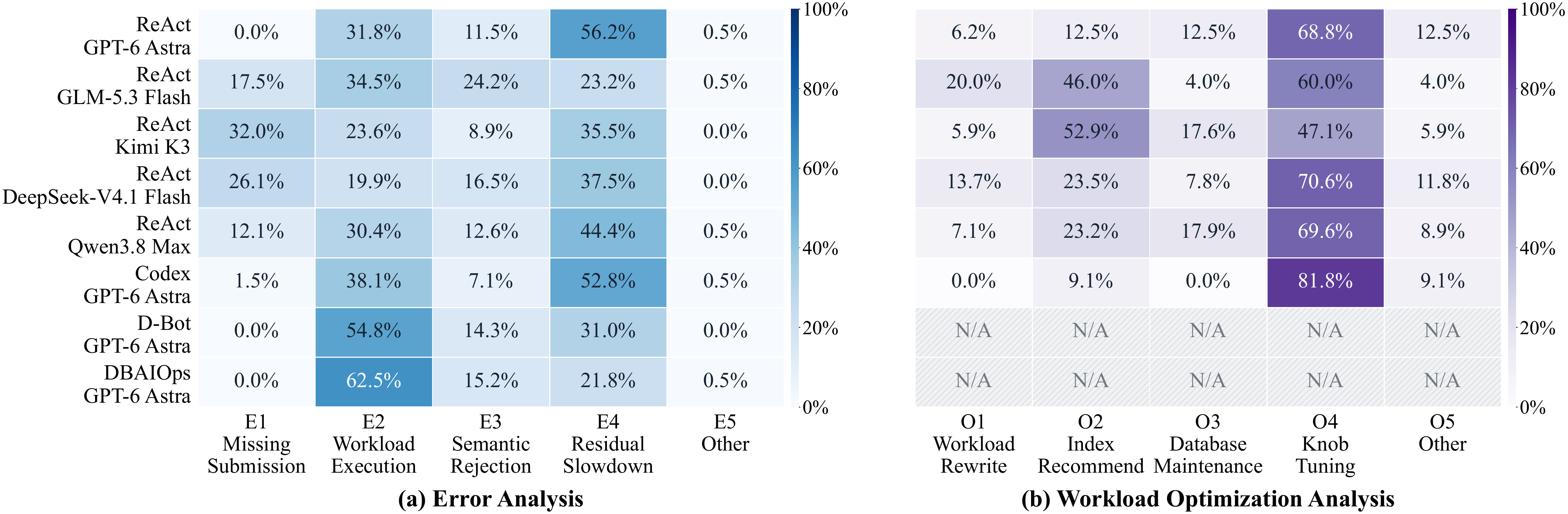}
    \caption{Detailed analysis of database diagnosis evaluations. Rows correspond to database agents, while columns denote error categories (left) and optimization categories (right). Note that O1--O4 may overlap because a sample can involve multiple optimization operations. D-Bot and DBAIOps are marked N/A because neither has eligible cases for this analysis.}
    \label{fig:detailed_analysis}
\end{figure*}
}

\newcommand{\figcase}{
\begin{figure*}[t]
    \centering
    \includegraphics[width=\textwidth]{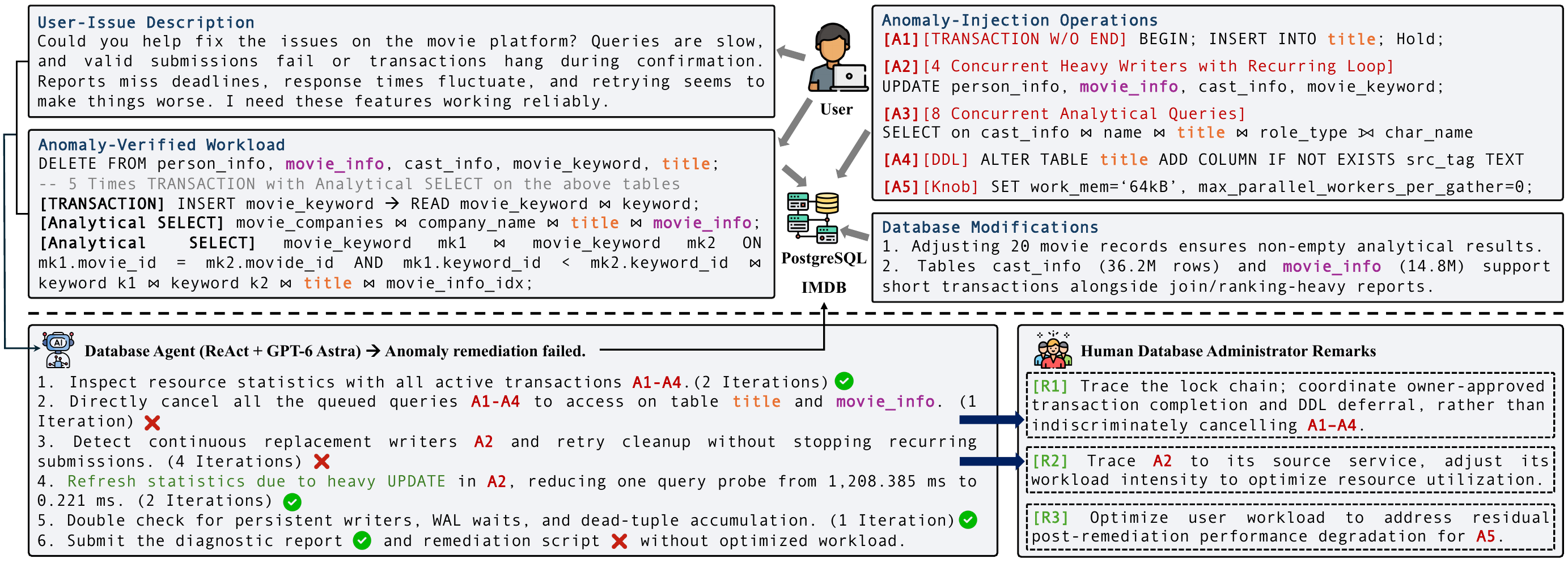}
    \caption{Case study of a ReAct agent with GPT-6 Astra. Following environment setup through database import, sample-specific modifications, and anomaly injection, database agent perform interactive diagnosis and remediation with user-issue description and anomaly-verified workload, ultimately outputting diagnostic report, remediation script, and optional optimized workload.}
    \label{fig:case_study}
\end{figure*}
}


\section{Introduction}
\label{sec:intro}

Database diagnosis aims to identify the root causes of anomalous behavior in database management systems (DBMSs) through systematic analysis of monitoring metrics, logs, and query executions, thereby guiding targeted remediation of the underlying issues.

\vpara{Prior Work and Remaining Limitations.}
Traditional rule-based or machine-learning-based approaches~\cite{dbsherlock, iSQUAD, AutoMonitor} have focused on detecting anomalies in time-series metrics and identifying their root-cause categories. DBPA~\cite{dbpa} supports the evaluation of such approaches by providing pre-collected metric data paired with anomaly labels, alongside fixed procedures for reproducing nine anomaly types.
With the emergence of large language models (LLMs), LLM-based systems~\cite{dbot, dbaiops, panda, gaussmaster} have extended diagnostic capabilities beyond root-cause categorization, enabling LLMs to emulate database administrators (DBAs) by interacting with DBMSs and generating comprehensive natural-language diagnostic reports that combine causal-chain analysis with remediation recommendations.

However, unlike conventional classification tasks that support training and evaluation on static datasets, evaluating interactive database agents requires live DBMS environments that reproduce the target anomalies. Moreover, time-series metric records collected through routine monitoring can be readily assembled into static training datasets, whereas constructing complete, executable anomalous DBMS environments at scale for interactive database agent training remains challenging.
D-Bot~\cite{dbot} releases 539 anomaly cases for PostgreSQL with detailed scenario descriptions, alongside 2,819 static training records.
More recently, DBA-Bench~\cite{dbabench} comprises 241 cases across PostgreSQL and TiDB for supporting interactive database agent evaluation.
Nevertheless, it contain only a few hundred evaluation cases without separate training data.

\vpara{Our Proposal.}
To bridge this gap, we introduce \bench, a large-scale benchmark for evaluating interactive database agents on anomaly diagnosis, remediation, and workload performance optimization. Specifically, \sbench distinguishes itself from existing benchmarks through three core strengths:
\textbf{(1) Scenario Diversity}, encompassing diverse anomaly-injection mechanisms and operational settings beyond fixed templates; 
\textbf{(2) Large-Scale Dataset}, comprising thousands of executable anomaly samples; 
and \textbf{(3) Real-World Grounding}, with anomaly samples derived from source material collected in real-world database scenarios.

\vpara{Challenges and Solutions.}
Constructing a large-scale benchmark that enables database agents to perform interactive diagnosis and remediation within live DBMS environments while capturing diverse, realistic anomaly scenarios presents the following key challenges:

\textbf{(1) How can \sbench ensure the diversity and realism of its anomaly samples?}
Addressing the limited availability of executable samples for training and evaluating interactive database agents requires expanding both dataset scale and the diversity of anomaly definitions.
To meet this demand, we collect source material for anomaly samples through two channels: 
(i) we obtain DBMS operational manuals from experienced DBAs and summarize common anomaly definitions, enriched with error codes and timeout information retrieved from official DBMS documentation, to establish a public source; 
and (ii) we collect real-world enterprise records of online diagnosis incidents as a private source.

\textbf{(2) How can different anomaly sources be transformed into complete, independently executable database anomaly samples at scale?}
Constructing complete, executable database anomaly samples from above anomaly definitions and incident records at scale is non-trivial, as this requires maintaining sample quality and diversity while ensuring that the specific scenarios reflect real-world database operations. To address this challenge, we propose an LLM-based database anomaly sample construction framework with two complementary pipelines tailored to the respective sources: 
(i) \textit{Manual-Guided Anomaly Synthesis}, which draws on official DBMS documentation and DBA manuals to sequentially synthesize user requests, database modification operations, and anomaly-injection scripts, respectively; 
and (ii) \textit{Incident-Guided Anomaly Reproduction}, which reconstructs complete database anomaly samples end to end from real-world enterprise online diagnosis incident records.

\textbf{(3) How can we reliably evaluate database diagnosis agents based on the simulative anomaly samples?}
LLM-based synthesis may introduce hallucinations that lead to validity issues such as (i) inconsistencies between user requests and anomaly-injection scripts (e.g., reporting table corruption or duplicate-key errors when the injection script only drops secondary indexes and holds locks, without reproducing either reported issue), and (ii) failures to reproduce the intended anomalous state (e.g., expecting row-level locks to block \texttt{SELECT} queries on a table while observing that all such queries complete successfully). 
These validity issues compromise evaluation reliability, making it difficult to determine whether an agent's failure stems from defects in the anomaly sample or limitations of the agent itself. 
To this end, we filter the constructed samples for anomaly-injection validity, and select challenging candidates based on reference database agent's diagnostic and remediation performance. Following iterative expert review and refinement, 454 expert-annotated samples are retained as a separate held-out test set for evaluating database diagnosis agents.
Furthermore, strong diagnostic reasoning does not necessarily translate into effective remediation that directly modifies the anomalous environment. We therefore evaluate database diagnosis agents using two complementary scoring components that assess both anomaly remediation outcomes and diagnostic report quality:
(i) \textit{Execution-Grounded Remediation Evaluation} scores remediation outcomes through comparing workload performance between the normal baseline and post-remediation states, along with an LLM-as-a-judge assessment of whether any agent-optimized workload preserves the identical semantics of the original anomaly-verified workload; 
and (ii) \textit{LLM-Judged Diagnostic Report Evaluation} uses the original anomaly sample as the ground-truth reference for LLM-based scoring across four dimensions: root-cause accuracy, evidence-grounded reasoning, remediation appropriateness, and operational safety.

\vpara{\bench.}
Building on these solutions, we construct \bench, comprising 6,940 synthesized training samples and a separate held-out test set of 454 expert-annotated samples, which spans three hardware configurations, two DBMSs, and three database schemas.
Unlike conventional benchmarks for text-to-SQL~\cite{omnisql, spider2} and question answering~\cite{squad}, database diagnosis and anomaly remediation require agents to inspect and modify the state of an anomalous database. We therefore package each anomaly sample as a self-contained sandbox providing an independent live DBMS environment that supports system monitoring, interactive diagnosis, anomaly remediation, and workload execution. Each sample in \sbench pairs this sandbox with a user request comprising an issue description and an anomaly-verified workload for evaluation.

\vpara{Contributions.}
Our main contributions are as follows:

\begin{itemize}[leftmargin=1em]
\setlength\itemsep{0em}
    \item \textbf{Benchmark for Interactive Database Agent Evaluation.}
    We introduce \bench, a benchmark spanning three hardware configurations, two DBMSs, and three database schemas, with 6,940 synthesized training samples and a held-out test set of 454 expert-annotated samples. Each anomaly sample pairs a user request—comprising an issue description and the workload under which the issue was observed—with an independent live DBMS environment that supports performance monitoring, interactive diagnosis, anomaly remediation, and workload execution.
    \item \textbf{Database Anomaly Sample Construction Framework.} We propose an LLM-based framework with two complementary pipelines to construct diverse anomaly samples at scale: \textit{Manual-Guided Anomaly Synthesis} follows a three-stage process grounded in official DBMS documentation and DBA manuals, while \textit{Incident-Guided Anomaly Reproduction} reconstructs samples end to end from real-world enterprise online diagnosis incident records.
    \item \textbf{Comprehensive Experimental Evaluation.} We conduct extensive evaluations of multiple database agents to assess their diagnostic and remediation capabilities. Even the best-performing agent, built on the ReAct framework and powered by GPT-6 Astra, achieves only 56.24, while most other agents score below 45, underscoring the challenges posed by \bench.
\end{itemize}
\section{Preliminary}
\label{sec:preliminary}

\figoverview

\subsection{Problem Definition}
\label{sec:problem_def}

DBAs typically address incoming online diagnosis requests through examining relevant performance metrics, identifying potential root causes, and performing timely remediation, ultimately summarizing this process in a diagnostic report with feasible remediation actions.
Hence, we formally define the database diagnosis task as follows:

\vspace{0.05in}\noindent \textit{Definition 1 (Database Diagnosis).} 
Upon receiving an online diagnosis request, the DBA diagnoses and remediates the issue based on the live anomalous DBMS environment \(E_\text{anomalous}\), user-issue description \(Iss_\text{user}\), and anomaly-verified workload \(W_\text{verify}\) under which the issue was observed, ultimately restoring---and potentially improving---the workload's expected execution performance and yielding a diagnostic report \(R_\text{diag}\) with remediation actions \(A_\text{reme}\).

\begin{equation}
\begin{aligned}
    DBA(E_\text{anomalous}, Iss_\text{user}, W_\text{verify}) \rightarrow (E_\text{normal}, R_\text{diag}, A_\text{reme}, W_\text{opt}).
\end{aligned}
\end{equation}

Here, \(DBA(\cdot)\) denotes a human DBA or an LLM agent acting in that role. 
Successful remediation restores the DBMS environment to a normal or optimized state \(E_\text{normal}\), with a potentially optimized workload \(W_\text{opt}\). To this end, we identify the fundermantal components of an anomaly sample for database diagnosis as follows:

\vspace{0.05in}\noindent \textit{Definition 2 (Database Anomaly Sample).}
Each sample \(S_i\) comprises a user request \(Req_i\) and the corresponding DBMS environment \(E_i\).

\begin{equation}
\begin{aligned}
    S_i &= (Req_i(Iss_{\text{user},i}, W_{\text{verify},i}), E_i(H_i, D_i, Ano_i)). \\
\end{aligned}
\end{equation}

Here, the user request \(Req_i\) comprises a user-issue description \(Iss_{\text{user},i}\) and an anomaly-verified workload \(W_{\text{verify},i}\), while the DBMS environment \(E_i\) includes the hardware configuration \(H_i\), database instance \(D_i\), and anomaly-triggering mechanism \(Ano_i\).


\subsection{ReAct Framework}
\label{sec:react}

ReAct~\cite{react}, which underpins our LLM-based database anomaly sample construction framework detailed in subsequent sections, interleaves natural-language reasoning with task-specific actions through a \textit{Thought--Action--Observation} loop. 
Starting from task-specific instructions, the LLM reasons about the next step (\textit{Thought}), invokes an external tool or interacts with the environment (\textit{Action}), and receives the resulting feedback (\textit{Observation}).
The accumulated reasoning traces, actions, and observations provide context for subsequent decisions, enabling the LLM to assess intermediate outcomes, revise its plan, and adjust its actions until it submits its final output or a predefined termination condition is met.
By coupling reasoning with execution feedback, ReAct supports adaptive task execution grounded in observations from external environments.


\section{Overview}
\label{sec:overview}
Database diagnosis and anomaly remediation require direct interaction with live DBMS environments to restore normal operation. 
We therefore encapsulate each sample in an isolated Docker~\cite{docker} container that enables users to instantiate sample environment on compatible hosts without relying on a fixed evaluation server.

Building on the preceding definitions and containerized environments, we propose an LLM-based framework to construct database anomaly samples for \sbench at scale, as illustrated in Figure~\ref{fig:overview}. Tailored to different anomaly sources, this framework offers two complementary pipelines to support diverse sample construction while allowing users to select suitable approach for their construction requirements. Given specified hardware configurations and database instances, these pipelines generate user requests and database operations required to instantiate anomalous environments:

\begin{itemize}[leftmargin=1em]
    \item \textbf{Manual-Guided Anomaly Synthesis} (Section~\ref{sec:ano_syn}) instantiates abstract anomaly definitions derived from DBA manuals and official DBMS documentation as executable samples through three sequential stages: (i) inferring user-observable symptoms from each anomaly definition to synthesize a user request comprising a user-issue description and an anomaly-verified workload; (ii) generating optional database modifications to support valid anomaly-verified workload execution and establish anomaly-specific data distributions; and (iii) constructing an anomaly-injection script to trigger the intended anomalous behavior.
    \item \textbf{Incident-Guided Anomaly Reproduction} (Section~\ref{sec:ano_rep}) reconstructs executable samples end to end from real-world enterprise online diagnosis records detailing user requests, root-cause analyses, and diagnostic traces. Compared with abstract anomaly definitions, these records provide richer, more explicit guidance, thereby supporting a single-stage, multi-turn process that directly generates and refines interdependent sample components using recorded evidence to mitigate hallucination risk.
\end{itemize}

Each anomaly sample comprises (i) a user request containing a user-issue description and an anomaly-verified workload; (ii) an anomalous DBMS container specified by its environmental configuration, database modification operations, and an anomaly-injection script; and (iii) an anomaly definition, as detailed in Section~\ref{sec:diagdbench}.
\section{Manual-Guided Anomaly Synthesis}
\label{sec:ano_syn}

We propose the \textit{Manual-Guided Anomaly Synthesis} pipeline to construct diverse, challenging anomaly samples from abstract definitions derived from DBA manuals and official DBMS documentation. It comprises three sequential stages for synthesizing user requests, database modifications, and anomaly injections, respectively.

\subsection{Anomaly Sample Source}
\label{sec:ano_def_source}

\begin{table}[t]
    \caption{17 single-anomaly root-cause definitions summarized from DBA operational manuals.}
    \label{tab:dba_anomaly_def}
    \centering
    \vspace{-1em}
    \begin{adjustbox}{width=0.45\textwidth}
    \begin{tabular}{@{}p{0.28\linewidth}p{0.68\linewidth}@{}}
        \toprule
        \textbf{Anomaly Layer} & \textbf{Root-Cause Definition} \\
        \midrule

        \multirow{4}{*}{\shortstack[l]{SQL Layer}}
        & SQL Workload Logic Anomaly \\
        & Schema and Index Design Anomaly \\
        & Optimizer Statistics Anomaly \\
        & Query Plan Selection Anomaly \\
        \midrule

        \multirow{8}{*}{\shortstack[l]{DBMS Layer}}
        & Transaction Lock Contention Anomaly \\
        & Multi-Version Concurrency Control (MVCC) Version Maintenance Anomaly \\
        & DBMS Configuration Anomaly \\
        & Database Memory Allocation Anomaly \\
        & Buffer Cache and Latch Contention Anomaly \\
        & Transaction Log Subsystem Anomaly \\
        & Replication Synchronization Anomaly \\
        & Data Block and Index Integrity Anomaly \\
        \midrule

        \multirow{5}{*}{\shortstack[l]{Server Layer}}
        & Storage I/O Throughput Anomaly \\
        & Storage Space Capacity Anomaly \\
        & CPU Resource Anomaly \\
        & Operating System Resource Limit Anomaly \\
        & Network and Connectivity Resource Anomaly \\

        \bottomrule
    \end{tabular}
    \end{adjustbox}
\end{table}

Explicit root-cause definitions grounded in real-world operational experience are essential for instantiating concrete database anomaly samples.
To this end, we collect DBMS operational manuals in which enterprise DBAs summarize their practical experience, and extract 17 single-anomaly root-cause definitions as presented in Table~\ref{tab:dba_anomaly_def}. 
To account for anomalies with multiple root causes, we enumerate combinations of these definitions 
(e.g., \textit{Schema and Index Design Anomaly + Transaction Lock Contention Anomaly}, where missing indexes can cause \texttt{UPDATE} statements to scan more rows and hold locks longer, causing blocking chains that delay other transactions and intensify lock contention),
yielding 833 general root-cause definitions for synthesizing more diverse and challenging scenarios.

Since distinct diagnostic paths may converge on the same root cause, a given root-cause definition can correspond to diverse user-observable symptoms (e.g., error messages or execution timeouts).
To enrich these definitions into more concrete anomaly samples, we build a knowledge base of user-observable symptoms through collecting error-code explanations and timeout reason analyses separately from the official documentation of PostgreSQL\footnote{\url{https://www.postgresql.org/docs/16/}.} and MySQL\footnote{\url{https://dev.mysql.com/doc/refman/8.4/en/}.}, as presented in Table~\ref{tab:dbms_anomaly_def}. 
Each entry is structured into a symptom name, official description, associated root-cause analysis and category, and effective remediation recommendations.

\begin{table*}[t]
    \caption{User-observable symptoms collected from the official website of each DBMS.}
    \label{tab:dbms_anomaly_def}
    \centering
    \vspace{-1em}
    \begin{adjustbox}{width=\textwidth}
    \begin{tabular}{@{}lccl@{}}
        \toprule
        \textbf{DBMS}
        & \textbf{\# Error Code Types}
        & \textbf{\# Timeout Types}
        & \textbf{Examples} \\
        \midrule

        PostgreSQL
        & 249
        & 13
        & \begin{tabular}[c]{@{}l@{}}
            Error Code: \texttt{40 P01 (deadlock\_detected)}, indicating a transaction rollback caused by a detected deadlock. \\
            Timeout: \texttt{statement\_timeout} aborts statements exceeding the time limit (0 disables the timeout). \\
          \end{tabular} \\
        \midrule

        MySQL
        & 6068
        & 17
        & \begin{tabular}[c]{@{}l@{}}    
            Error Code: \texttt{1213 (ER\_LOCK\_DEADLOCK)} indicates a deadlock during lock acquisition; the transaction should be retried. \\    
            Timeout: \texttt{innodb\_lock\_wait\_timeout} sets the maximum row-lock wait time for an InnoDB transaction.
          \end{tabular} \\
        \bottomrule
    \end{tabular}
    \end{adjustbox}
\end{table*}


After constructing the knowledge base, we retrieve relevant entries using formatted anomaly root-cause definitions as queries. Our hybrid lexical retrieval approach comprises two components:
\begin{itemize}[leftmargin=1em]
    \item \textbf{Field-weighted BM25}~\cite{bm25} prioritizes key-field matches (e.g., symptom names and official descriptions) while accounting for term rarity, frequency, and entry length to mitigate bias toward long or repetitive entries.
    \item \textbf{Term overlap} quantifies both query-term coverage and the proportion of entry terms shared with the query.
\end{itemize}

For each DBMS, we independently retrieve entries from its error-code and timeout knowledge bases. Let $K$ denote either knowledge base, $Q$ the distinct terms extracted from the root-cause definition, and $T_d$ the distinct terms in the retrieval fields $\mathcal{F}$ of entry $d \in K$ (e.g., symptom name and official description). 
We aggregate term counts as $\operatorname{tf}_w(t,d)=\sum_{f \in \mathcal{F}}w_f\operatorname{tf}(t,d_f)$, where $d_f$ is the field content, $\operatorname{tf}(t,d_f)$ counts occurrences of term $t$, and $w_f$ is a predefined field weight fixed during retrieval. 
With the weighted entry length $L_w(d)=\sum_{t \in T_d}\operatorname{tf}_w(t,d)$ and its average $\overline{L}_w$ over $K$, the field-weighted BM25 score is:
\begin{equation}
    s_{\mathrm{BM25}}(Q,d)
    =
    \sum_{t \in Q \cap T_d}
    \operatorname{IDF}(t)
    \frac{(k_1+1)\operatorname{tf}_w(t,d)}
    {\operatorname{tf}_w(t,d)+k_1\left(1-b+bL_w(d)/\overline{L}_w\right)},
    \label{eq:weighted_bm25}
\end{equation}
where $k_1=1.5$, $b=0.75$, and $\operatorname{IDF}(t)=\ln\!\left(1+\frac{|K|-n_t+0.5}{n_t+0.5}\right)$, with $n_t$ counting entries whose retrieval fields contain $t$.

We normalize BM25 within $K$ as $\widehat{s}_{\mathrm{BM25}}(Q,d)=s_{\mathrm{BM25}}(Q,d)/M_Q$, where $M_Q=\max_{d' \in K}s_{\mathrm{BM25}}(Q,d')$, assigning zero when $M_Q=0$. 

The term-overlap score combines query-term coverage with the proportion of entry terms matching the query:
\begin{equation}
    s_{\mathrm{overlap}}(Q,d)
    =
    0.7 \cdot \frac{|Q \cap T_d|}{|Q|}
    +
    0.3 \cdot \frac{|Q \cap T_d|}{|T_d|},
    \label{eq:term_overlap}
\end{equation}
with a score of zero if either term set is empty.

We compute the final relevance score as a weighted combination of the normalized BM25 and term-overlap scores as follows:
\begin{equation}
    s(Q,d)
    =
    0.7\,\widehat{s}_{\mathrm{BM25}}(Q,d)
    +
    0.3\,s_{\mathrm{overlap}}(Q,d).
    \label{eq:retrieval_score}
\end{equation}

For each root-cause definition, we retain the highest-scoring entries with nonzero relevance scores from each knowledge base $K$ (top 25\% for error codes and 40\% for timeouts).
To enhance the diversity of final anomaly definitions, we select up to two entries from each retained set, prioritizing less frequently retrieved entries and favoring higher relevance scores when retrieval counts are equal.
Each retrieved symptom is paired with its corresponding root-cause definition, yielding 1,666 complete anomaly definitions per DBMS for subsequent anomaly sample construction.
Notably, user-defined anomaly definitions can also serve as alternative inputs without modifying the subsequent three-stage synthesis procedure.


\subsection{User Request Synthesis}
\label{sec:user_issue}
Given an anomaly definition, we first synthesize a user request comprising a user-issue description and an anomaly-verified workload under which the anomaly is observed. 
In practice, users may encounter unexpected performance degradation or execution errors, thereby submitting online requests to DBAs for anomaly diagnosis and remediation. 
To simulate this process, we adopt the ReAct framework to guide an LLM in iteratively generating the workload and issue description from the anomaly definition.
Specifically, the prompt for user request synthesis comprises three components:
\begin{itemize}[leftmargin=1em]
    \item \textbf{Task Overview} instructs the LLM to reason step by step and generate the workload that exposes the specified anomaly, together with the corresponding user-issue description.
    \item \textbf{Environment-Reset Skill} guides the LLM in restoring the database to its initialized state when needed, preventing earlier changes from affecting subsequent validation. 
    \item \textbf{Sample-Relevant Information} specifies the hardware configuration (CPU, memory, etc.), the DBMS (PostgreSQL~\cite{pg} or MySQL~\cite{mysql}), the database instance and associated schema (TPC-H~\cite{tpch}, TPC-DS~\cite{tpcds}, or IMDB~\cite{job}), and the anomaly definition.
\end{itemize}

In particular, ``Task Overview'' specifies three constraints on workload generation to extend database agent evaluation beyond diagnosis and remediation to include performance optimization:
\begin{itemize}[leftmargin=1em]
    \item \textbf{Workload execution time} should exceed one minute, preferably through complex analytical queries, to provide scope for workload execution-time optimization.
    \item \textbf{Workload execution consistency} requires the workload to produce consistent results across repeated executions, enabling equivalence checks between the original user workload and the workload optimized by database agents.
    \item \textbf{Workload performance optimization} could be feasible through techniques such as index recommendation, SQL rewriting, etc.
\end{itemize}

Guided by this prompt, the LLM can choose between the following two actions during iterative reasoning to generate and refine the anomaly-verified workload and user-issue description:
\begin{itemize}[leftmargin=1em]
    \item \textbf{Verify Action} leverages a workload verifier to execute the candidate workload repeatedly, collect execution results and latencies, and determine whether the workload satisfies the above constraints. The above verifications are returned as observations to guide subsequent refinement and action selection.
    \item \textbf{Finish Action} terminates the synthesis process and outputs the finalized anomaly-verified workload and user-issue description once the observations satisfy the verification criteria.
\end{itemize}

\subsection{Database Modification Synthesis}
\label{sec:data_modify}
To assess whether database agents can resolve anomalies without making additional unnecessary changes to the database or altering the original semantics of the user request, we require the \texttt{SELECT} queries in the anomaly-verified workload to return non-empty results, supporting workload consistency assessment based on comparisons of pre- and post-remediation execution results.
Moreover, some anomaly scenarios require specific data distributions (e.g., concentrating orders on a single buyer so that concurrent payment transactions compete for locks on the same rows), thereby necessitating targeted modifications to the given database instances as well as enhancing their diversities across anomaly samples.

To this end, we synthesize database modification operations through iterative reasoning and interaction within the ReAct framework, with its prompt consisting of the following components:
\begin{itemize}[leftmargin=1em]
    \item \textbf{Task Overview} guides the LLM's step-by-step reasoning and action selection for synthesizing database modification operations.
    \item \textbf{Environment-Reset Skill and Sample-Relevant Information} remain unchanged from their counterparts in Section~\ref{sec:user_issue}.
    \item \textbf{User Request} provides the user-issue description and associated anomaly-verified workload synthesized in Section~\ref{sec:user_issue}.
\end{itemize}

Guided by the prompt and available observations, the LLM selects one of the following two actions at each reasoning iteration:
\begin{itemize}[leftmargin=1em]
    \item \textbf{Query Action} takes one of two forms: (i) executing queries to inspect relevant data characteristics of the target database instance, such as whether any rows satisfy a specified predicate; or (ii) executing candidate modification workloads to assess their feasibility. In both cases, the execution results are returned as observations to guide subsequent reasoning and action selection.
    \item \textbf{Finish Action} finalizes and outputs the database modification operations once accumulated observations indicate that the instance satisfies the target conditions. Additionally, the LLM also summarizes the updated database characteristics most relevant to the anomaly definition and user request in natural language, forming part of the input to the anomaly-injection stage (e.g., identifying two hotspot rows accessed by the workload to guide the construction of a concurrent transaction that acquires their locks in reverse order to reproduce deadlock-induced failures).
\end{itemize}

\subsection{Anomaly Injection Synthesis}
\label{sec:ano_inject}
After synthesizing the database modification operations, we proceed to the final and most critical stage: anomaly-injection script synthesis. 
As in the preceding stages, we utilize a ReAct-based LLM to construct and refine the script through successive rounds of reasoning and interaction. To guide this process, the prompt for synthesizing anomaly-injection scripts is structured as follows:
\begin{itemize}[leftmargin=1em]
    \item \textbf{Task Overview} guides the LLM in constructing and validating an anomaly-injection script through step-by-step reasoning and action selection. Notably, since the self-contained environment supports operating-system-level operations that may be not occurred in real-world scenarios, we restrict anomaly injection to DBMS-level operations that simulate business logic.
    \item \textbf{Environment-Reset Skill, Sample-Relevant Information, and User Request} retain the same content as in Section~\ref{sec:data_modify}.
    \item \textbf{Database Feature Summary} describes the updated database's sample-relevant data distributions presented in Section~\ref{sec:data_modify}.
\end{itemize}

Guided by the prompt, the LLM iteratively constructs the anomaly-injection script through selecting one of the following two actions:
\begin{itemize}[leftmargin=1em]
    \item \textbf{Evaluation Action} submits the candidate anomaly-injection script to an anomaly evaluator, which evaluates the script following three steps: (i) executing the user workload in the initialized self-contained environment and recording its baseline performance; (ii) running the script to inject the anomaly; and (iii) rerunning the same user workload to check for performance degradation or execution errors, with the evaluation logs returned as observations to guide subsequent script refinement.
    \item \textbf{Finish Action} returns the final anomaly-injection script when the collected evidence indicates that the target anomaly has been successfully induced, resulting in observable performance degradation or execution errors in the anomaly-verified workload.
\end{itemize}

\section{Incident-Guided Anomaly Reproduction}
\label{sec:ano_rep}

To enhance the realism and source diversity of \sbench, we propose the \textit{Incident-Guided Anomaly Reproduction} framework, which reconstructs complete anomaly sample end to end from real-world online diagnosis records collected within the enterprise.

\begin{table}[t]
    \caption{Field definitions of real-world online diagnosis incident records collected from enterprise.}
    \label{tab:oncall_char}
    \centering
    \vspace{-1em}
    \footnotesize
    \begin{adjustbox}{width=0.48\textwidth}
    \begin{tabular}{@{}p{0.24\linewidth}p{0.62\linewidth}@{}}
        \toprule
        \textbf{Field Name} & \textbf{Field Description} \\
        \midrule

        Incident Name & Brief title of the user request. \\

        Severity Level & Operational severity of the reported issue (e.g., critical, high, medium, or low). \\

        Responsible Party & Optional information identifying who is accountable for the incident. \\
        
        Surface Symptom & An operation- or service-level label derived from the user request, such as Data Manipulation Language (DML), Data Definition Language (DDL), or resource management, rather than a specific failure symptom. \\
        
        Root Cause & A rule-derived candidate anomaly category (e.g., workload execution, resource bottlenecks, or DBMS configuration), rather than a verified root cause. \\

        Diagnostic Evidence & Specific evidence collected while addressing the user request, comprising a key message, its interpretation, and the method used to localize the issue. \\
        
        Category Analysis & Free-text description compiled from incident summaries, messages, and diagnostic evidence to support root-cause category assignment. \\
        
        Incident Summary & A structured overview of the incident, involving observed symptoms, affected users, investigation progress, and the reported conclusion. \\
        
        \bottomrule
    \end{tabular}
    \end{adjustbox}
\end{table}

\subsection{Anomaly Sample Source}
\label{sec:ano_rep_source}

We collected conversation records of online diagnosis incidents submitted to DBAs in enterprise environments during May--July 2026 as source material for anomaly sample reconstruction, with the record structure detailed in Table~\ref{tab:oncall_char}. Alongside operational anomalies in DBMSs, the collected records contain requests outside the scope of database diagnosis, such as front-end platform issues, authorization requests, etc. Therefore, we applied the following filters, retaining 1,086 records for incident-guided anomaly reproduction:
\begin{itemize}[leftmargin=1em]
    \item \textbf{Knowledge-Inquiry Filtering} excludes those incidents whose surface symptom field is explicitly marked as an inquiry.
    \item \textbf{Platform-Issue Filtering} removes the records assigned to the platform-related category in root cause field, such as platform functionality, user interface interaction, and infrastructure issues outside the scope of DBMS operation and diagnosis.
    \item \textbf{Instance-Life-Cycle Filtering} excludes records whose root cause category identifies instance-management operations, such as instance creation, capacity expansion, or database migration.
    \item \textbf{Information-Sufficiency Filtering} removes records lacking key attributes or sufficient descriptive detail to identify their specific database anomaly category.
\end{itemize}

\subsection{Single-Stage Anomaly Reproduction}
\label{sec:allinone}

Each online diagnosis incident record provides a detailed, concrete description of an anomaly sample, supplying explicit context that reduces reliance on speculative inference and helps mitigate hallucination during LLM inference. 
Building on the preceding three-stage synthesis pipeline, we adopt a simplified, single-stage process to jointly construct and refine interdependent sample components through multi-turn interaction, thereby accelerating construction while maintaining consistency with the recorded incident.

To guide the LLM in interpreting the collected incidents and constructing complete anomaly samples consistent with the source records, we incorporate additional skills tailor to the incident record into the input prompt, which comprises the following components:
\begin{itemize}[leftmargin=1em]
    \item \textbf{Task Overview} guides the LLM through step-by-step reasoning to construct a complete anomaly sample based on the incident.
    \item \textbf{Environment-Reset Skill} remains the same as in Section~\ref{sec:user_issue}.
    \item \textbf{User-Request Interpretation Skill} clarifies record fields and maps reported issues to common anomaly definitions. It also specifies the operational scope of self-contained environment: lock contention, query plan degradation, and database resource pressure are reproducible, whereas platform- and infrastructure-level incidents (e.g., production network failures and resource scheduling faults) can only be approximated through database-visible effects. Reported symptoms for evaluator check are classified as performance degradation, blocking timeout, or execution error.
    \item \textbf{Executable Scenario Construction Skill} guides the LLM in converting real-world online diagnosis incident records into standardized, executable, and measurable anomaly sample components, and provides detailed instructions for anomaly validation.
    \item \textbf{Reproduction Protocol Skill} defines the output specifications and the final validation criteria. Each sample must pass the evaluator checks derived from the online diagnosis record and include all five components: an anomaly scenario definition, a user-issue description, an anomaly-verified workload, database modification operations, and an anomaly-injection script.
    \item \textbf{Sample-Relevant Information} contains the hardware configuration, DBMS, database instance and its schema, and specific incident record with relevant components described in Table~\ref{tab:oncall_char}.
\end{itemize}

Anomaly reproduction primarily follows the multi-turn ReAct framework described in Section~\ref{sec:ano_inject}, in which the LLM selects between the ``Evaluation Action'' and ``Finish Action''. When the LLM selects the ``Evaluation Action'', the anomaly evaluator validates the candidate sample and returns execution logs as observations to guide subsequent refinement of the relevant output components. This process continues until the LLM chooses the ``Finish Action''.

\section{\sbench for Database Diagnosis}
\label{sec:diagdbench}
Leveraging this anomaly construction framework, we establish \sbench for interactive evaluation of database agents on anomaly diagnosis, remediation, and workload performance optimization.

\subsection{Benchmark Preparation}
\label{sec:bench_details}

Building on 3,332 anomaly definitions and 1,086 online diagnosis records, we utilize the framework described in Sections~\ref{sec:ano_syn} and~\ref{sec:ano_rep} constructing 8,836 candidate anomaly samples across different DBMSs at scale.
To promote linguistic diversity, we employ four LLMs for synthesis: GPT-5.5~\cite{gpt55}, o3~\cite{o3}, GLM-5.2~\cite{glm52}, and DeepSeek-V4-Pro~\cite{deepseekv4}.
Meanwhile, we allow up to 20 iterations per construction stage while retaining the default setting of the underlying LLM.

Although an evaluator checks anomaly-injection success during sample construction, its observations primarily guide subsequent LLM reasoning rather than mandatory instructions to further refinement after failure.
Moreover, we observed that, for some samples, the LLM reached the iteration limit without issuing the ``Finish Action'', leaving synthesis completion unconfirmed.
Therefore, we filter the constructed samples using anomaly evaluator presented in Section~\ref{sec:ano_inject}, excluding those whose anomaly-verified workloads exhibit no substantial performance change after injection.
Following anomaly validity filtering, we retain 7,394 valid samples in total. 


From this pool, we curate challenging candidates for expert review to establish a high-quality, independent held-out test set for database agent evaluation.
For each sample, we first measure the anomaly-verified workload's normal-operation runtime $t_0$, then employ an o3-based ReAct database agent to diagnose and remediate the injected anomaly, after which we rerun the workload and record its execution time $t_1$. 
Failed executions and timeouts are assigned $t_1 = \infty$. 
We rank samples in descending order of $r = t_1 / t_0$ that prioritizes greater residual performance degradation or failure to restore successful workload execution, selecting the top 600 candidates across DBMSs and construction pipelines for expert annotation (Section~\ref{sec:human_anno}).
Expert review and refinement yield 454 held-out test samples, leaving 6,940 samples available for training.

For public release of the held-out test set, we package anomaly-injection scripts in a protected form to reduce the risk of agents identifying root causes through code inspection rather than database diagnosis. Meanwhile, the released artifacts preserve executable entry points compatible with the original evaluation workflow without directly exposing human-readable injection source code.

\subsection{Expert Review}
\label{sec:human_anno}

Following Section~\ref{sec:bench_details}, we select 600 anomaly samples as candidates for a held-out test set to support unified database agent evaluation. Despite validity checks and difficulty-based agent selection, these LLM-generated samples may still differ from real-world incidents (e.g., claiming disk-full errors without reproducing storage exhaustion, or login failures when the injected anomaly merely slows down queries). We therefore invite a DBA with five years of experience diagnosing and resolving user-reported production database issues to review the candidates from four perspectives:
\begin{itemize}[leftmargin=1em]
    \item \textbf{Anomaly Scenario Reasonability} assesses whether each operation reflects business purposes, maintenance requirements, or plausible operational mistakes within a realistic scenario.
    \item \textbf{Trigger Mechanism Correctness} checks whether anomaly injections conform to the target DBMS's execution rules without relying on operating-system-level interventions.
    \item \textbf{Script Executability and Configuration Effectiveness} assesses whether the script executes successfully with its intended effects. Specifically, the DBA inspects logs and database state for execution errors and verifies that configuration changes apply to the intended connections throughout workload execution. 
    \item \textbf{Consistency Between Anomaly Implementation and User Request} checks whether the reported symptoms match the injection operations and observed behavior. Unsupported claims, such as stating with database deletion when none occurred or query timeouts with completed execution, are corrected or removed.
\end{itemize}

Beyond these checks, the DBA may propose optional refinements that increase diagnostic difficulty based on the original anomaly injection mechanism, prioritizing related concurrent operations rather than simply increasing workload intensity or waiting times. 

For the 600 selected anomaly samples, together with their corresponding anomaly-verification logs and collected system metrics, the DBA may approve samples satisfying the above criteria, discard low-quality samples, or recommend revisions for those worth retaining, organizing any revision feedback into three components:
 
\begin{itemize}[leftmargin=1em]
    \item \textbf{Anomaly Double Check} records unresolved validity concerns and required verifications, such as checking whether an anomaly-injection timeout due to insufficient execution budget was incorrectly treated as evidence of successful anomaly reproduction.
    \item \textbf{Anomaly Modification} identifies operations that violate the above criteria and suggests corresponding targeted revisions.
    \item \textbf{Anomaly Refinement} specifies additional anomaly injections, with supporting explanations, to increase diagnostic difficulty.
\end{itemize}

Guided by this, we leverage a ReAct agent powered by GPT-5.5 to repair and refine the anomaly samples, incorporating further DBA feedback iteratively until approval without additional changes.


\begin{table}[t]
    \caption{Overview statistics of \bench. Workload, Data Mod., and Anomaly Inj. report statement or operation counts (min / max / avg) for anomaly-verified workloads, database modifications, and anomaly-injection scripts.}
    \label{tab:bench_stat}
    \centering
    \vspace{-1em}
    
    \begin{adjustbox}{width=0.46\textwidth}
    \begin{tabular}{@{}llrrrr@{}}
        \toprule
        DBMS & Split & Workload & Data Mod. & Anomaly Inj. & Samples \\
        \midrule
        \multirow{2}{*}{PostgreSQL}
            & Train & $1 / 450 / 44.29$ & $0 / 118 / 1.65$
            & $1 / 373 / 23.43$ & $3{,}644$ \\
            & Test  & $2 / 278 / 45.73$ & $0 / 11 / 1.71$
            & $1 / 107 / 27.31$ & $236$ \\
        \midrule
        \multirow{2}{*}{MySQL}
            & Train & $1 / 314 / 31.29$ & $0 / 57 / 0.95$
            & $0 / 181 / 14.85$ & $3{,}296$ \\
            & Test  & $2 / 156 / 19.19$ & $0 / 18 / 1.75$
            & $2 / 64 / 18.33$ & $218$ \\
        \bottomrule
    \end{tabular}
    \end{adjustbox}
\end{table}


\subsection{Statistical Analysis}
\label{sec:data_statistic}
\sbench comprises 6,940 synthesized training samples and a separate held-out test set of 454 expert-annotated samples, with the specific components of each sample as follows:
\begin{itemize}[leftmargin=1em]
    \item \textbf{Anomaly Definition} provides a structured description of the target anomaly, including root-cause definition, observable symptoms, and injection success criteria where available.
    \item \textbf{Environment Configuration} specifies hardware configuration, DBMS, database instance, and database connection information.
    \item \textbf{User-Issue Description} describes the user-observed symptoms in natural language during workload execution.
    \item \textbf{Anomaly-Verified Workload} denotes the SQL workload used to observe anomaly-injection effects and assess database agents in anomaly remediation and workload performance optimization.
    \item \textbf{Database Modification Operations} comprise SQL operations that establish sample-specific data and schema prerequisites, with an empty file when no modifications are specified.
    \item \textbf{Anomaly-Injection Script} is an executable script for establishing the target anomaly in the configured DBMS environment.
\end{itemize}

To promote environmental diversity, we construct anomaly samples for PostgreSQL and MySQL using three hardware configurations: (i) 4 CPUs, 4 GB RAM, a combined RAM-and-swap limit of 8 GB, and 64 GB storage; (ii) 8 CPUs, 8 GB RAM, a combined RAM-and-swap limit of 16 GB, and 128 GB storage; and (iii) 16 CPUs, 16 GB RAM, a combined RAM-and-swap limit of 32 GB, and 256 GB storage. 
Meanwhile, we import prebuilt database instances based on different schemas of TPC-H~\cite{tpch}, TPC-DS~\cite{tpcds}, and IMDB~\cite{job} to provide the initial data.
Additionally, \sbench instantiates 7,394 distinct anomaly definitions from the sources in Sections~\ref{sec:ano_def_source} and~\ref{sec:ano_rep_source} to enhance scenario diversity.
Table~\ref{tab:bench_stat} illustrates observable variation in the numbers of SQL statements in anomaly-verified workloads and database modifications, as well as operations in anomaly-injection scripts, highlighting structural diversity across samples beyond differences in their execution environments.

Compared with existing datasets and benchmarks, as presented in Table~\ref{tab:benchmark_comparison}, \sbench provides thousands of executable anomaly samples across diverse execution environments, enabling interactive database diagnosis, anomaly remediation and workload performance optimization while providing a foundation for training database agents to improve both capabilities. 
Notably, except the samples included in \bench, we provide an LLM-based anomaly sample construction framework with two complementary pipelines that enables users to generate customized samples from their own anomaly specifications. This framework supports the construction of simulated scenarios tailored to specific production environments and operational requirements, facilitating the iterative development and improvement of database agents.



\subsection{Evaluation Metrics}
\label{sec:eval_metrics}

\begin{table*}[t]
    \centering
    \begin{minipage}{\textwidth}
        \centering
        \caption{Comparison of database diagnosis datasets. Note: ``WPO Eval. Support'' denotes workload performance optimization evaluation support; ``Training and Test Sources'' reports anomaly-definition counts. ``-'' indicates unreported counts or inapplicable settings; ``*'' marks static samples, such as time-series metrics or diagnostic records for LLM supervised fine-tuning.}
        \label{tab:benchmark_comparison}
        \scriptsize
        \vspace{-1em}

        \renewcommand{\arraystretch}{1}
        \setlength{\tabcolsep}{5pt}

        \resizebox{\linewidth}{!}{%
            \begin{tabular}{@{}lccccccc@{}}
                \toprule
                \textbf{Benchmark}
                & \textbf{\# Hardware Configs}
                & \textbf{\# DBMSs}
                & \textbf{\# Schemas}
                & \textbf{\# Training Sources}
                & \textbf{\# Test Sources}
                & \textbf{Interactive Operation Support}
                & \textbf{WPO Eval. Support} \\
                \midrule
                
                DBPA~\cite{dbpa}
                & 4 & 2 & 4 & 9$^{*}$ & 9$^{*}$ & No & No \\

                D-Bot~\cite{dbot}
                & - & 1 & 3 & 2,819$^{*}$ & 539
                & No & No \\
                
                DBA-Bench~\cite{dbabench}
                & - & 2 & - & - & 241 
                & Yes & No \\

                \midrule

                \bench
                & 3 & 2 & 3 & 6,940 & 454 & Yes & Yes \\

                \bottomrule
            \end{tabular}%
        }
    \end{minipage}
\end{table*}


Since \sbench primarily targets interactive database anomaly remediation while also assessing diagnostic quality, we adopt two complementary evaluation protocols: \textit{Execution-Grounded Remediation Evaluation} and \textit{LLM-Judged Diagnostic Report Evaluation}. 
Specifically, we conduct standardized evaluation and scoring of the diagnostic reports, remediation scripts, and optional optimized anomaly-verified workloads submitted by database agents upon completion of the database diagnosis and anomaly remediation.


\subsubsection{Execution-Grounded Remediation Evaluation}
For each sample $i$, we initialize a container according to its environment configuration and execute the original anomaly-verified workload to record the baseline results $O_i^{\mathrm{0}}$ and latency $t_{0,i}$. Then, we inject the anomaly and execute the agent-provided remediation script with DBA privileges. In the post-remediation environment, we execute the agent-provided optimized anomaly-verified workload, if available, or the original workload otherwise, recording results $O_i^{\mathrm{1}}$ and latency $t_{1,i}$. To this end, we define the remediation latency ratio as:
\begin{equation}
    r_i = \frac{t_{1,i}}{t_{0,i}}.
    \label{eq:recovery_latency_ratio}
\end{equation}

Since we encourage workload optimization during diagnosis and remediation, it is necessary to assess whether each optimized workload preserves the identical intent and observable semantics of the original user request.
Comparing execution results alone cannot establish semantic equivalence, while volatile values (e.g., timestamps, random values, or generated identifiers) may introduce spurious mismatches. We therefore use an LLM judge~\cite{llmasajudge,llmasajudge1} to assess semantic equivalence based on the original and optimized workloads together with their execution results $O_i^{0}$ and $O_i^{1}$. 
Specifically, the judge evaluates both workloads as complete, ordered sequences of statements, including query outputs, ordering requirements, and transaction state. Equivalence claims lacking sufficient evidence are rejected, and we define the binary semantic-equivalence indicator $C_i=1$ when the judge accepts equivalence and $C_i=0$ otherwise.

Finally, we evaluate anomaly remediation in terms of execution success, workload semantic equivalence, and post-remediation performance. A binary execution-validity indicator $E_i$ equals 1 if the post-remediation workload completes without errors and its latency is less than eight times the normal baseline latency ($r_i<8$), and 0 otherwise. Using $E_i$, $C_i$, and $r_i$, we then compute the anomaly remediation score for each sample on a scale of 0 to 100 as follows:
\begin{equation}
    R_i =
    \begin{cases}
        0, & C_i=0 \text{ or } E_i=0,\\
        80 - 10(r_i-1), & C_i=E_i=1,\ 1 \leq r_i < 8,\\
        80 + 20(1-r_i), & C_i=E_i=1,\ 0.2 \leq r_i < 1,\\
        100, & C_i=E_i=1,\ 0 \leq r_i < 0.2.
    \end{cases}
    \label{eq:fault_recovery_score}
\end{equation}
Conditional on $C_i=E_i=1$, matching the normal baseline latency yields 80 points. Higher scores indicate improved workload performance, whereas lower scores reflect residual slowdown. A score of zero is assigned when workload execution encounters errors or post-remediation latency is at least eight times the baseline.

\subsubsection{LLM-Judged Diagnostic Report Evaluation}
\label{sec:eval_diag_report}
An LLM judge evaluates each diagnostic report against the original anomaly sample, comprising its ground-truth anomaly definition, user-issue description, anomaly-verified workload, and anomaly-injection script, scoring each of the following four dimensions on a 0--100 scale:
\begin{itemize}[leftmargin=1em]
    \item \textbf{Root-Cause Accuracy ($S_{\mathrm{rca}}$)} assesses whether the diagnostic report accurately identifies the injected root causes and reconstructs their corresponding causal chain.
    \item \textbf{Evidence-Grounded Reasoning ($S_{\mathrm{reason}}$)} assesses whether the report provides rigorous, internally consistent reasoning supported through observed symptoms and available evidence.
    \item \textbf{Remediation Appropriateness ($S_{\mathrm{remediation}}$)} assesses whether the proposed actions are technically executable and address the anomaly without unnecessary operations.
    \item \textbf{Operational Safety ($S_{\mathrm{safe}}$)} assesses whether the actions are confined to the scope of the user request, thereby avoiding unintended harm to unrelated data or database operations.
\end{itemize}


For each sample $i$, the diagnostic report score aggregates the four dimension scores using the following weights:
\begin{equation}
    D_i =
        0.45S_{\mathrm{rca},i}
        + 0.25S_{\mathrm{reason},i}
        + 0.20S_{\mathrm{remediation},i}
        + 0.10S_{\mathrm{safe},i}.
    \label{eq:diagnosis_score}
\end{equation}

\subsubsection{Overall Score}
\label{sec:eval_overall_score}
For each sample $i$, we compute the overall score as a weighted sum of the remediation and diagnostic report scores, prioritizing anomaly remediation effectiveness:
\begin{equation}
    S_i = 0.70R_i + 0.30D_i.
    \label{eq:sample_score}
\end{equation}

For each method $m$, we compute its final score by averaging the sample-level scores over the evaluated subset $\mathcal{T}_m$ of the test set:
\begin{equation}
    \operatorname{Score}(m)
    =
    \frac{1}{|\mathcal{T}_m|}
    \sum_{i \in \mathcal{T}_m}
    \left(
        0.70R_{m,i} + 0.30D_{m,i}
    \right).
    \label{eq:method_score}
\end{equation}

\section{Experiments}
\label{sec:exp}
In this section, we presents a detailed evaluation of different database agents on the 454-sample held-out test set of \bench.

\subsection{Experimental Setup}
\label{sec:exp_setup}

\begin{table*}[t]
    \centering
    \begin{minipage}{\textwidth}
        \centering
        \caption{Evaluation scores of eight database agents on \bench. Each cell reports the mean score over the corresponding samples, using the agent-proposed optimized workload when available and the original anomaly-verified workload otherwise. D-Bot's Overall score includes only the 236 PostgreSQL samples, as it does not support MySQL (--).}
        \label{tab:main_result}
        \scriptsize
        \vspace{-1em}
        \renewcommand{\arraystretch}{1.1}
        \setlength{\tabcolsep}{5pt}
        \resizebox{\linewidth}{!}{%
            \begin{tabular}{@{}c*{8}{c}@{}}
                \toprule
                \multirow{2}{*}[-\dimexpr(\aboverulesep+\belowrulesep+\cmidrulewidth)/2\relax]{\textbf{\bench}} & \multicolumn{5}{c}{\textbf{ReAct}} & \textbf{Codex} & \textbf{D-Bot} & \textbf{DBAIOps} \\
                \cmidrule(lr){2-6}\cmidrule(lr){7-7}\cmidrule(lr){8-8}\cmidrule(lr){9-9}
                & GPT-6 Astra & GLM-5.3 Flash & Kimi K3 & DeepSeek-V4.1 Flash & Qwen3.8 Max & GPT-6 Astra & GPT-6 Astra & GPT-6 Astra \\
                \midrule
                \textbf{PostgreSQL} & 56.81 & 41.43 & 39.40 & 42.25 & 43.64 & 53.62 & 20.09 & 23.27 \\
                \textbf{MySQL} & 55.61 & 44.58 & 46.01 & 43.98 & 46.31 & 51.22 & -- & 11.03 \\
                \midrule
                \textbf{Overall} & 56.24 & 42.97 & 42.65 & 43.10 & 44.92 & 52.44 & 20.09 & 20.55 \\
                \bottomrule
            \end{tabular}%
        }
    \end{minipage}
\end{table*}

\subsubsection{Baselines}
We provide all database agents with a standardized prompt template comprising the following three components:
\begin{itemize}[leftmargin=1em]
    \item \textbf{Task Overview} guides step-by-step LLM reasoning and action selection for database diagnosis and anomaly remediation. It also encourages performance optimization through semantics-preserving rewrites of the anomaly-verified workload or tuning operations incorporated into the remediation code.
    \item \textbf{Sample-Relevant Information} specifies hardware configuration, database instance, DBA credentials, and the user request with an issue description and an anomaly-verified workload.
    \item \textbf{Output Format} requires database agents to submit a diagnostic report, remediation code, and an optional optimized workload.
\end{itemize}

Specifically, we evaluate eight LLM-based database agents while retaining the default hyperparameter settings of their underlying LLMs, and the architectures of these agents are described as follows:

\vpara{ReAct-based Database Agents}
We instantiate five database agents using the ReAct~\cite{react} framework with GPT-6 Astra~\cite{gpt6astra}, Qwen3.8-Max~\cite{qwen38max}, DeepSeek-V4.1-Flash~\cite{deepseekv41flash}, Kimi K3~\cite{kimik3}, and GLM-5.3-Flash~\cite{glm53flash} as the underlying models. Each database agent is limited to 20 interactions per sample, and at each reasoning iteration, the LLM can choose between the following two actions:
\begin{itemize}[leftmargin=1em]
    \item \textbf{Code Action} generates a Bash script that is executed within the sample container for database inspection, anomaly remediation, or workload validation, with the execution output returned as an observation to guide subsequent reasoning and action selection.
    \item \textbf{Finish Action} concludes the interaction with a diagnostic report, reusable remediation code, and an optional optimized workload.
\end{itemize}

\vpara{Codex-based Database Agent}
We evaluate the Codex~\cite{codex} scaffold with GPT-6 Astra using its native agent loop and Model Context Protocol (MCP) tools. Specifically, we supplement the standardized inputs with tool definitions and interaction instructions: \textit{db\_execute} executes diagnostic and remediation scripts within the sample container and returns observations; \textit{db\_finish\_diagnosis} submits the diagnostic report, remediation code, and optional optimized workload. Each sample allows at most 20 database-tool actions.

\vpara{D-Bot~\cite{dbot}}
We evaluate D-Bot with GPT-6 Astra for PostgreSQL diagnosis using its native multi-expert workflow, which dynamically assigns specialized agents to retrieve monitoring tools and expert knowledge, investigate candidate root causes through tree-search reasoning and reflection, and cross-review findings before generating final report. We supplement standardized inputs with 12 monitoring snapshots collected over time from the sample environment, such as resource utilization, database statistics, and blocking relationships. Since native D-Bot produces diagnostic findings and recommendations rather than executable remediation code, a separate LLM-based adapter generates a remediation script and an optional optimized anomaly-verified workload from its finalized report. We retain the implemented search settings: two simulations per expert and a maximum depth of 24 steps per search chain.

\vpara{DBAIOps~\cite{dbaiops}}
We evaluate DBAIOps with GPT-6 Astra using its implemented knowledge-based question-answering workflow. Given standardized inputs and a read-only sample-environment snapshot, DBAIOps queries an external knowledge-base service and augments the LLM input with retrieved material to generate a diagnostic report without iterative observation or remediation. A separate LLM call converts the finalized report into a remediation script and an optional optimized workload, preserving the native diagnostic workflow. Bounded retries handle retrieval and model-request failures, incomplete responses, and invalid artifact formats without constituting additional database diagnosis rounds.

\subsubsection{Evaluation Settings}
\label{sec:exp_eval_setting}
Database agents are evaluated according to the scoring procedures in Section~\ref{sec:eval_metrics}. Specifically, each workload evaluation comprises one warm-up run followed by five timed runs. The median latencies before anomaly injection and after remediation define as $t_{0,i}$ and $t_{1,i}$, from which we compute the remediation latency ratio $r_i$. 
We use GPT-6.1 Sol~\cite{gpt61sol} as the judge for both workload semantic-equivalence assessment and diagnostic report evaluation. For the former, we obtain three valid binary judgments $c_i^{(j)} \in \{0,1\}$ through separate evaluations and determine $C_i$ by majority voting~\cite{majorvoting0, majorvoting1}. For the latter, we conduct three scoring rounds and define $D_i$ as the median of the resulting weighted scores.

\subsection{Main Results}
\label{sec:exp_main_res}
Table~\ref{tab:main_result} presents assessment results for eight database agents on \bench, with the key findings as follows:

\figanalysis

\textbf{\sbench remains challenging even for frontier LLMs paired with agent scaffolds.} 
Experimental results demonstrate that the ReAct-based database agent with GPT-6 Astra achieves the highest overall score among the eight database agents, reaching 56.24 out of 100. Apart from GPT-6 Astra paired with ReAct or Codex, other approaches score below 47 on each supported DBMS, highlighting the challenges of accurate diagnosis, effective remediation, and workload performance optimization on \bench.

\textbf{Database-specific diagnosis approaches achieve degraded evaluation scores on \bench.} 
D-Bot and DBAIOps employ knowledge-augmented diagnostic workflows primarily designed for diagnostic reports generation. Although we use an LLM to translate these reports into remediation scripts, D-Bot scores 20.09 on its supported PostgreSQL samples, while DBAIOps achieves 20.55 on the full test set.
These observed performance suggests that directly translating diagnostic reports into effective remediation actions remains challenging, further motivating the significance of interactive remediation that enables database agents to incorporate execution feedback, refine their actions, and verify recovery.

\textbf{Database diagnosis and remediation performance varies across LLM backbones and agent scaffolds.}
With the ReAct framework, input prompts, and available tools held fixed, GPT-6 Astra outperforms the second-best model, Qwen3.8 Max, by 11.32 points in overall score. Meanwhile, with GPT-6 Astra as the fixed underlying model, ReAct achieves a higher overall score than Codex, reaching 56.24 compared with 52.44. 
These results underscore the importance of both the underlying LLM and the surrounding agent harness in enhancing the capability of database agent.

\textbf{Stronger performance on specific DBMS does not necessarily translate into an advantage on others.} 
Within the ReAct scaffold, Kimi K3 scores 39.40 on PostgreSQL and 46.01 on MySQL, whereas DeepSeek-V4.1 Flash achieves 42.25 and 43.98, respectively. This reversal in their relative ranking demonstrates that overall scores alone do not fully capture the models' distinct strengths across anomaly scenarios instantiated on different DBMSs.

\subsection{Detailed Analysis}
\label{sec:exp_detailed_analysis}
We exclude successful recoveries with remediation latency ratios $r_i \in [0.8, 1.2]$ and analyze the remaining samples in Figure~\ref{fig:detailed_analysis}. Specifically, we report error-category proportions among failed recoveries and successful recoveries with $r_i > 1.2$, alongside optimization-operation prevalence among successful recoveries with $r_i < 0.8$.

\subsubsection{Error Analysis}
\label{sec:exp_error}
Each error case is assigned one of the following outcome-based labels: \textit{Missing Submission} (E1; failure to submit expected and complete outcomes within the maximum iteration), \textit{Workload Execution} (E2; anomaly-verified workload execution fails or times out), \textit{Semantic Rejection} (E3; the semantic judge rejects equivalence between the original and optimized workloads), \textit{Residual Slowdown} (E4; execution and functional validation succeed, but performance remains degraded with $r_i > 1.2$), or \textit{Other} (E5).
Figure~\ref{fig:detailed_analysis}(a) illustrates the following recurring failure patterns:

\begin{itemize}[leftmargin=1em]
    \item \textbf{Workload Execution Bottleneck.}
    Workload execution failures constitute the most or second-most prevalent error category for most database agents, highlighting the persistent challenge of effective remediation for the anomaly samples in \bench.

    \item \textbf{Incomplete Performance Recovery.}
    Residual Slowdown accounts for 56.2\% and 52.8\% for ReAct and Codex with GPT-6 Astra, respectively, with substantial proportions also observed across other agents. Despite explicit instructions to optimize the execution performance of the anomaly-verified workload, database agents typically focus on addressing the primary root-cause actions without sufficiently validating end-to-end performance, thereby leaving residual performance bottlenecks unresolved.

    \item \textbf{Challenges in semantics-preserving workload rewriting.}
    Beyond limitations in diagnostic capabilities, some agents struggle in producing optimized workload rewrites that satisfy semantic-equivalence validation, thereby resulting in unexpected failures even after the primary anomaly has been resolved.
\end{itemize}

\subsubsection{Workload Optimization Analysis}
\label{sec:exp_opt}
Among successful recoveries with $r_i < 0.8$, we identify five categories of performance optimization as follows: \textit{Workload Rewrite} (O1; workload modifications accepted as semantically equivalent), \textit{Index Recommend} (O2), \textit{Database Maintenance} (O3; statistics updates and table or index maintenance), \textit{Knob Tuning} (O4), and \textit{Other} (O5). Figure~\ref{fig:detailed_analysis}(b) reveals the characteristics of the optimization behavior as follows:

\begin{itemize}[leftmargin=1em]
    \item \textbf{Diverse Optimization Strategies.}
    Database agents adopt various strategies for workload performance optimization, with an average speedup of $4.42\times$, thereby further mitigating residual slowdowns beyond simply restoring workload functionality.
    \item \textbf{Dominant Tuning Operations.}
    Knob tuning is the most prevalent operation across most database agents, appearing in 47.1\%--81.8\% of these samples, while index recommendation accounts for 9.1\%--52.9\%. These results suggest that frontier-LLM-based database agents can leverage conventional database configuration and indexing knowledge for workload performance optimization.
    \item \textbf{Potential for Semantics-Preserving Rewriting.}
    Compared with failed samples due to semantic-equivalence violation, successful cases satisfy this requirement while achieving substantial latency reductions, thereby motivating further improvements in database agents' ability to enhance execution efficiency while reliably preserving the original workload's semantics.
\end{itemize}

\subsection{Case Study}
\label{sec:case_supp}

Figure~\ref{fig:case_study} illustrates the interactive database diagnosis and remediation of an anomaly in an IMDB-based PostgreSQL environment using the ReAct agent powered by GPT-6 Astra.

\vpara{Anomaly Sample.}
In the PostgreSQL database \texttt{IMDB}, supporting movie-content management and analytics, an injected long-lived transaction and recurring background workloads cause blocking and performance degradation. Specifically, users encounter workload-wide slowdowns and transaction stalls when executing a mixed workload combining five write-and-confirm transactions with four analytical queries over large shared tables. These disruptions stem from an uncommitted transaction blocking DDL, thereby queuing subsequent readers, alongside four recurring bulk-update clients and eight analytical clients operating under restrictive database knobs. Together, these conditions combine lock-induced blocking, constrained query execution, and sustained resource contention between foreground transactions and background analytics.

\figcase

\vpara{Database Diagnosis.}
Across ten iterations, the agent followed a conventional DBA workflow by inspecting locks, performing query and session cleanup, and refreshing statistics, yielding modest performance gains. Despite identifying recurring updates, four iterations of session-level diagnosis and backend cleanup failed to manage upstream submissions, leaving remediation incomplete.

Beyond this failure, the ``Human Database Administrator Remarks'' in Figure~\ref{fig:case_study} highlight three operational shortcomings:
(i) terminating active sessions risks disrupting ongoing work, whereas transaction completion and DDL execution should be coordinated with workload owners (R1);
(ii) repeatedly removing replacement connections addresses transient symptoms rather than tracing recurring updates to their submitting source and regulating workload intensity according to resource demand (R2);
and (iii) when unexpected slowdowns persist, DBAs routinely inspect DBMS knobs and concurrent resource usage to address residual bottlenecks rather than treating symptom removal as sufficient recovery (R3).

\vpara{Discussion.}
In practice, production DBMSs routinely support numerous concurrent sessions, requiring DBAs to resolve user-visible disruptions while coordinating ongoing workloads and maintaining system-wide performance. However, existing database agents struggle to diagnose complex anomalies and coordinate interdependent sessions while preserving scheduling dependencies and using shared resources efficiently. Even after successful anomaly remediation, they tend to finish diagnosis once the superficial blocking is resolved, without systematically checking for residual performance bottlenecks.
Reliable database agents must therefore integrate business-aware workload coordination with system-wide performance optimization to restore functionality and expected performance rather than merely alleviate symptoms, which motivates further research into developing and refining these capabilities.

\section{Related Work}
\nocite{*}

\subsection{Database Anomaly Benchmarks}
DBPA~\cite{dbpa} defines nine common anomalies and reproduction procedures, collects monitoring metrics, and synthesizes compound anomalies to construct training and test sets for root cause classification. 
More recently, DBA-Bench~\cite{dbabench} provides 241 production-fidelity diagnosis samples for PostgreSQL and TiDB, evaluating database agents through multi-turn interactions with live databases, but offers limited samples without separate training data. In contrast, \sbench provides 454 held-out evaluation samples and thousands of training samples across diverse environmental settings.

\subsection{Database Diagnosis}

\vpara{Traditional Methods}
Early studies predominantly formulate database diagnosis into root cause classification through rule-based or machine-learning-based methods. DBSherlock~\cite{dbsherlock} maps structured anomaly predicates extracted from monitoring metrics to a DBA-informed causal graph, ranking candidate root causes and generating causal explanations through backward symptom tracing and rule-based pruning of infeasible paths. 
iSQUAD~\cite{iSQUAD} focuses on diagnosing intermittent slow queries (iSQ) in cloud databases through extracting typed key performance indicators anomaly patterns, then removing redundant dependencies, and grouping similar iSQs by type-oriented pattern integration clustering. A Bayesian case model is subsequently used to characterize each cluster, substantially reducing labeling overhead while enabling online root-cause diagnosis of incoming iSQs. 
AutoMonitor~\cite{AutoMonitor} identifies anomalous metric windows using Long Short-Term Memory (LSTM) reconstruction errors, detects metrics with significant distributional shifts via the Kolmogorov--Smirnov test, and matches similar historical anomalies using a weighted K-Nearest Neighbors (KNN) model.

\vpara{LLM-based Methods.}
D-Bot~\cite{dbot} integrates anomaly monitoring, profiling, diagnosis, and report generation within a multi-agent framework, with GaussMaster~\cite{gaussmaster} extending this paradigm to openGauss. Given anomaly metrics, it retrieves documentation and tools from knowledge base, combining tree-search reasoning, reflection, and cross-agent review for report summarization.
Panda~\cite{panda} supports performance debugging through global and instance-specific contextual grounding, evidence-based response verification, risk- and impact-aware action assessment, and feedback-driven historical case integration. 
DBAIOps~\cite{dbaiops} explores and expands diagnostic paths in heterogeneous knowledge graphs encoding DBA expertise, supplying retrieved reasoning chains and anomalous metrics to an LLM for root-cause analysis and remediation recommendations.

Despite the availability of D-Bot's public dataset, early LLM-based approaches lacked a unified benchmark for evaluating interactive diagnosis in live DBMSs. \sbench addresses this gap with diverse, executable samples and separate training and held-out evaluation sets, enabling unified assessment of diagnostic accuracy, remediation effectiveness, and optional workload optimization.
\section{Conclusion}

We introduce \sbench for training and evaluating database agents on interactive database diagnosis, anomaly remediation, and optional workload performance optimization. \sbench simulates diverse, realistic anomaly samples in self-contained, executable DBMS environments spanning three hardware configurations, two DBMSs, and three database schemas. To construct anomaly samples at scale, we propose an LLM-based framework with two complementary pipelines: \textit{Manual-Guided Anomaly Synthesis}, grounded in DBA manuals and official DBMS documentation, and \textit{Incident-Guided Anomaly Reproduction}, derived from real-world enterprise online diagnosis records. Built using these pipelines, \sbench comprises 6,940 synthesized training samples and a held-out test set of 454 expert-annotated samples. Experimental evaluations across multiple database agents reveal persistent performance limitations, underscoring the challenges posed by \bench.



\bibliographystyle{ACM-Reference-Format}
\bibliography{_reference}

\end{document}